\documentclass{article}
\usepackage{spconf,amsmath,graphicx}
\usepackage{url}            
\usepackage{booktabs}       
\usepackage{amssymb}
\usepackage[normalem]{ulem}
\useunder{\uline}{\ul}{}
\usepackage{bibspacing}

\newcommand\method{AVES}

\title{AVES: Animal Vocalization Encoder based on Self-Supervision}
\name{Masato Hagiwara}
\address{Earth Species Project}
\begin{document}
%
\maketitle
\begin{abstract}
The lack of annotated training data in bioacoustics hinders the use of large-scale neural network models trained in a supervised way. In order to leverage a large amount of unannotated audio data, we propose \method~(Animal Vocalization Encoder based on Self-Supervision), a self-supervised, transformer-based audio representation model for encoding animal vocalizations. We pretrain \method~on a diverse set of unannotated audio datasets and fine-tune them for downstream bioacoustics tasks. Comprehensive experiments with a suite of classification and detection tasks have shown that \method~outperforms all the strong baselines and even the supervised ``topline'' models trained on annotated audio classification datasets. The results also suggest that curating a small training subset related to downstream tasks is an efficient way to train high-quality audio representation models. We open-source our models\footnote{\url{https://github.com/earthspecies/aves}}.
\end{abstract}
\begin{keywords}
bioacoustics, self-supervision
\end{keywords}

\vspace{-0.5em}
\section{Introduction}
\vspace{-0.5em}

\begin{figure*}[t]
\begin{center}
\includegraphics[scale=0.35]{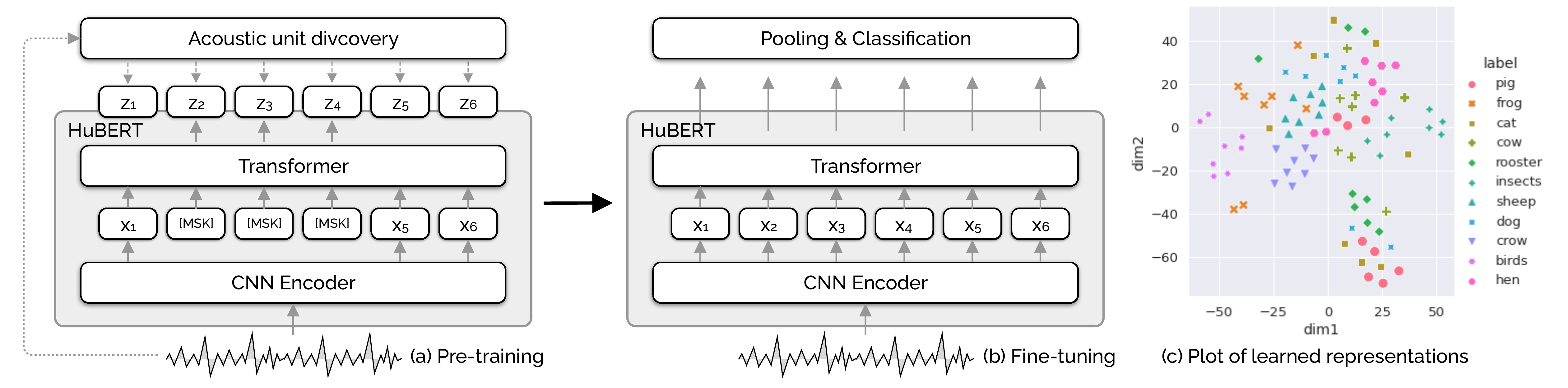} 
\vspace{-1.5em}
\caption{Overview of \method~(a) pretraining and (b) fine-tuning, and (c) t-SNE plot of learned representations}
\vspace{-1.5em}
\label{fig:overview}
\end{center}
\end{figure*}

The use of machine learning, especially deep neural network models, has become increasingly popular in bioacoustics in recent years, driven by the increased amount of bioacoustic data recorded by affordable recording devices~\cite{tuia2022perspectives}.

However, when trained in a supervised way, deep neural networks typically require a large amount of annotated data, and collecting high-quality annotation in bioacoustics requires deep expertise and is often costly in terms of the time required for manual labeling. This lack of labelled data in bioacoustics hinders the use of large-scale models and most recent models still rely on species-specific customized CNNs (convolutional neural networks)  trained on a small amount of task-specific annotated data~\cite{stowell2022computational}. Although a few studies~\cite{kahl2021birdnet,bain2021automated} pretrained from large datasets such as AudioSet~\cite{gemmeke2017audioset}, this requires a large annotated dataset of general sound.

One solution to this issue is self-supervision, a type of machine learning technique where the training signals come from the data itself in the form of pseudo labels. In recent years, large-scale self-supervised models and transfer learning have been hugely successful in related domains such as NLP~\cite{devlin2019bert,raffel2020exploring}, computer vision~\cite{grill2020bootstrap,chen2020big}, and human speech processing~\cite{baevski2020wav2vec2,chen2021wavlm}. There has been prior work on training general-domain audio representation models~\cite{turian2022hear,wang2022towards}, although those models are not designed specifically for bioacoustics and contain few datasets of animal sound. These findings lead us to expect that similar self-supervised approaches are also effective for bioacoustic tasks.

In this paper, we propose \method~(Animal Vocalization Encoder based on Self-supervision\footnote{Pronounced /ei vi:z/. ``Aves'' is the class name for birds, although our method supports diverse animal species, not just birds.}), a self-supervised, transformer-based audio representation model for encoding animal vocalizations for downstream bioacoustic tasks. For modeling raw waveforms with self-supervision, we use HuBERT (Hidden Unit BERT)~\cite{hsu2021hubert}, an audio representation model originally proposed for human speech. HuBERT obtained one of the top results in the SUPERB benchmark~\cite{yang2021superb}. We pretrain \method~on unannotated audio data that include not only animal vocalizations, but also other sounds (e.g., human speech and environmental sound).  We emphasize that the pretraining is purely self-supervised and does not rely on any annotation, besides choosing which instances to include. We evaluate \method~on a suite of classification and detection tasks, two of the most common tasks considered in the bioacoustic literature~\cite{stowell2022computational}. Our experiments show that \method~outperforms all the baselines (including pretrained ResNet and VGGish~\cite{hershey2017cnn} models pretrained on millions of videos), even the supervised topline models trained on annotated audio classification datasets.

Our contributions are as follows: with extensive benchmarking, we show that self-supervised methods originally proposed for human speech processing can be successfully applied to the domain of bioacoustics. This paper is also one of the few first studies~\cite{stowell2022computational} showing that, given appropriate pretraining, transformers work well for data-scarce domains such as bioacoustics, where CNNs are still by far the most common approach. We open source our pretrained models.

\vspace{-0.5em}
\section{Method}
\vspace{-0.5em}

\method~heavily relies on HuBERT~\cite{hsu2021hubert}, a self-supervised audio representation model originally proposed for human speech.

\vspace{-0.5em}
\subsection{HuBERT pretraining}
\vspace{-0.5em}

Unlike human written language where text can be used as self-supervision signals, human speech and bioacoustics data are usually given in terms of continuous, raw waveforms, where discrete units such as phonemes, syllables and words are not available a priori. In order to bootstrap optimization objectives from raw continuous audio, HuBERT leverages acoustic unit discovery, which finds discrete sound representation tokens $Z^{(i)} = [z^{(i)}_1, ..., z^{(i)}_T]$, for instance $i$ from raw sound data as noisy target labels, where $T$ is the number of frames per instance. Specifically, in the first stage of HuBERT training, cluster labels are obtained by applying $k$-means clustering on 39-dimensional MFCC features.

The HuBERT architecture (Fig.~\ref{fig:overview} left) consists of a CNN encoder that converts raw waveform into per-frame continuous audio representations (at a rate of 50 frames per second), and a transformer encoder which takes a masked sequence of audio representations and produces hidden representations. These hidden representations are then used to predict the acoustic units. The model is pretrained on a BERT-like masked language objective by minimizing the pretraining loss ${\cal L}_{\rm pt}$, which captures how well the model is able to predict the acoustic units at the masked positions:

$$
{\cal L}_{\rm pt} = -\sum_i \sum_{t \in M} \log p(z_t^{(i)} | \tilde X^{(i)})
$$
where $M$ is the set of integer masked positions and $\tilde X^{(i)}$ is the masked input for instance $i$. Applying the loss only on the masked positions $M$ forces the model to learn the acoustic and long-range representations from unmasked positions. The distribution $p(\cdot)$ is parameterized as a softmax on the similarities between the projected model output $\{{\bf h}_t\} = f(\tilde X)$ and an embedding vector ${\bf e}_c$ for class $c$:

$$
p(z_t | \tilde X) = {\rm softmax}_c\left( {\rm sim}({\bf h}_t {\bf W}, {\bf e}_c) / \tau \right)
$$
where ${\bf W}$ is a projection matrix, $\tau$ is the temperature parameter, and ${\rm sim}(\cdot, \cdot)$ denotes the consine similarity.

In order to further improve the quality of the discovered acoustic units, after the initial HuBERT model has been pretrained, the clustering is repeated on the features extracted from some internal layer (6th for {\tt base}, 12th for {\tt large} architectures) of the pretrained model itself. The $k$-means algorithm is run on those features and used for the second stage of training. One could also repeat this process more than twice, although we used the model of the second iteration.

\vspace{-0.5em}
\subsection{Transfer learning to bioacoustics}
\vspace{-0.5em}

We apply the learned representations to two tasks: classification and detection. For classification, each audio instance is assigned a label from a set of predefined classes, such as individuals and species. In detection, subsections of interest in long recordings are identified, often along with their properties such as call types. We use a sliding window approach, as is common for detection tasks~\cite{dufourq2021automated}, where long recordings are
broken up into short (potentially overlapping) segments and the model makes multi-label predictions about whether each segment contains vocalizations (of potentially different classes). Thus, both tasks can be solved by the same ML model, with only slight modification to the classification layer, as we describe below.

After extracting the audio representations $\{{\bf h}_t\}$ from the pretrained HuBERT model, we apply mean-pooling to extract a single summary vector per each instance. We found in preliminary experiments that the use of mean-pooling worked better than, for example, taking the first token for embedding, as commonly done in BERT~\cite{devlin2019bert}. After the pooling, a fully-connected linear layer is applied to extract logit values for classes. We denote this classification layer (pooling and a linear layer) by $f_{\rm cls}$.



For classification, $f_{\rm cls}({\bf h}_t)$ is fed to a softmax layer, and the network is fine-tuned with a cross entropy loss. For detection, we use a multi-label classification objective for fine-tuning where the activations are fed to a sigmoid layer and the network is trained on a binary cross entropy loss. For both tasks, the entire network was optimized via gradient-based updates, except for the CNN encoder, whose weights are frozen during fine-tuning, as done in HuBERT~\cite{hsu2021hubert}.

\vspace{-0.5em}
\section{Experiments}
\vspace{-0.5em}

\subsection{Pretraining details}
\vspace{-0.5em}

\begin{table}[!t]
\begin{center}
{\small
\begin{tabular}{@{}llllllrr@{}}
\toprule
              & FSD           & AS          & \multicolumn{3}{l}{AudioSet+VGGSound} &    &  \\
 Config.      & 50k           & 20k          & bio        & nonbio     & all           &   \# segs. & hours      \\ \midrule
{\tt core}   & \checkmark       & \checkmark      &                  &                  &                     & 67k  & 153    \\
{\tt bio}    & \checkmark       & \checkmark      & \checkmark       &                  &                     & 142k  & 360   \\
{\tt nonbio} & \checkmark       & \checkmark      &                  & \checkmark        &                     & 142k & 360   \\
{\tt all}    & \checkmark       & \checkmark      &                  &                  & \checkmark          & 1846k & 5054 \\ \bottomrule
\end{tabular}
}
\end{center}
\vspace{-1.0em}
\caption{Dataset configurations used for pretraining}
\vspace{-1.0em}
\label{table:datasets}
\end{table}

\begin{table*}[!t]
\begin{center}
{\small
\begin{tabular}{@{}lllllllllllll@{}}
\toprule
       & \multicolumn{5}{l}{Classification}    & \multicolumn{5}{l}{Detection}          & \multicolumn{2}{l}{Auxiliary} \\ \midrule
       & wtkn  & bat   & cbi   & hbdb  & dogs  & dcase & enab  & hiceas & rfcx  & gib   & esc           & sc            \\ \midrule
lr       & 0.776          & 0.661          & 0.156          & 0.751          & 0.885          & 0.143          & 0.247          & 0.221          & 0.030          & 0.006          & 0.428          & 0.535           \\
svm      & {\ul 0.870}    & 0.720          & 0.139          & 0.779          & {\ul 0.914}    & 0.146          & 0.299          & 0.218          & 0.038          & 0.039          & 0.478          & 0.572           \\
dt       & 0.661          & 0.474          & 0.025          & 0.700          & 0.626          & 0.095          & 0.183          & 0.223          & 0.009          & 0.007          & 0.240          & 0.230           \\
gbdt     & 0.758          & 0.674          & 0.038          & 0.759          & 0.827          & 0.104          & 0.235          & 0.216          & 0.009          & 0.007          & 0.338          & 0.481           \\
xgb      & 0.808          & 0.692          & 0.097          & 0.772          & 0.842          & 0.124          & 0.270          & 0.214          & 0.012          & 0.007          & 0.403          & 0.525           \\ \midrule
rn18     & 0.752          & 0.443          & 0.357          & 0.697          & 0.662          & 0.161          & 0.325          & 0.280          & 0.064          & 0.164          & 0.500          & 0.926           \\
rn50     & 0.799          & 0.548          & 0.295          & 0.696          & 0.633          & 0.183          & 0.282          & 0.304          & 0.055          & 0.215          & 0.235          & 0.936           \\
rn152    & 0.743          & 0.483          & 0.330          & 0.645          & 0.511          & 0.154          & 0.280          & 0.255          & 0.069          & 0.248          & 0.365          & 0.929           \\
rn18p    & 0.735          & 0.532          & 0.509          & 0.649          & 0.705          & 0.223          & 0.462          & 0.262          & 0.079          & \textbf{0.316} & 0.590          & 0.936           \\
rn50p    & 0.735          & 0.560          & 0.548          & 0.673          & 0.763          & 0.178          & 0.424          & 0.284          & 0.087          & 0.155          & 0.545          & 0.946     \\
rn152p   & 0.720          & 0.544          & 0.573    & 0.662          & 0.741          & 0.198          & 0.429          & 0.273          & 0.085          & 0.230                & 0.540          & 0.946     \\
vggish   & 0.847          & {\ul 0.743}    & 0.440          & {\ul 0.808}    & 0.906          & {\ul 0.335}    & {\ul 0.535}    & {\ul 0.463}    & {\ul 0.140}    & 0.150          & 0.705          & 0.948     \\ \midrule
rn152p-s & 0.835          & 0.606          & {\ul 0.583}    & 0.700          & 0.799          & 0.300          & 0.520          & 0.326          & 0.097          & {\ul 0.303}    & \textbf{0.788} & 0.947     \\
vggish-s & 0.794          & 0.723          & 0.445          & 0.801          & 0.806          & 0.325          & 0.509          & 0.428          & \textbf{0.143} & 0.186          & 0.708          & {\ul 0.950}   \\ \midrule
\method-{\tt bio} & \textbf{0.879} & \textbf{0.748} & \textbf{0.598} & \textbf{0.810} & \textbf{0.950} & \textbf{0.392} & \textbf{0.555} & \textbf{0.629} & 0.130 & 0.284          & {\ul 0.773}    & \textbf{0.964}           \\ \bottomrule
\end{tabular}
}
\end{center}
\vspace{-1.0em}
\caption{Main results. We used accuracy for classification and auxiliary tasks, and mean average precision for detection tasks. The best and the second best metrics are highlighted and underlined per each dataset.}
\vspace{-1.0em}
\label{table:results}
\end{table*}

We pretrained \method~on combinations of publicly available audio datasets, namely, FSD50K~\cite{fonseca2020fsd50k}, AudioSet~\cite{gemmeke2017audioset}, and VGGSound~\cite{chen2020vggsound}. We used FSD50K and the balanced subset of AudioSet as the {\tt core} configuration for \method~(Table~\ref{table:datasets}, first row) and included it in all other configurations.

We also built the {\tt bio} pretraining configuration (Table~\ref{table:datasets}, second row), which consists of subsets of AudioSet and VGGSound that contain animal vocalizations. Specifically, we chose all audio segments that have corresponding labels under the {\tt Animal} (ID: {\tt /m/0jbk}) concept in the AudioSet ontology, and the {\tt animals} class group in VGGSound. In order to investigate the effect of domain shift between pretraining and fine-tuning, we also built a control dataset configuration called {\tt nonbio} (Table~\ref{table:datasets}, third row) which has the same size as {\tt bio} but consists of randomly selected segments taken from AudioSet and VGGSound.

Finally, the {\tt all} configuration includes all the segments in AudioSet and VGGSound. All configurations and their size (the number of segments and total length in hours) are shown in Table~\ref{table:datasets}\footnote{Neither AudioSet nor VGGSound distributes the audio segments and the actual data we used and the statistics are as of when we obtained the corresponding YouTube data in 2022}. Note that our pretraining datasets are relatively smaller than, for example, the YouTube dataset containing millions of videos used to pretrain VGGish.

For pretraining HuBERT models, we used 200 clusters (at both the first and the second stage), a learning rate of $2.0 \times 10^{-4}$, an effective batch size of 700 seconds (450 seconds for {\tt large}) of audio, and 100k training steps (150k steps for {\tt large}). We determined these hyperparameters on the basis of the results from preliminary experiments with a smaller subset of the AudioSet dataset. Other hyperparameters followed the original HuBERT {\tt base} (12-layer, 768-unit transformer) and {\tt large} (24-layer, 1024-unit transformer) configurations. We used the implementation from fairseq\footnote{\url{https://github.com/facebookresearch/fairseq}}.

The baseline models include logistic regression (LR), support vector machine (SVM), decision tree (DT), gradient-boosted decision tree (GBDT), XGoost (XGB, \cite{chen2016xgboost}), ResNet~\cite{he2016deep} (ResNet18, ResNet50, ResNet152, both random and pretrained weights on ImageNet), and VGGish~\cite{hershey2017cnn}, a VGG-like architecture~\cite{simonyan2015very} pretrained on a large YouTube audio dataset. We followed the training settings from BEANS~\cite{hagiwara2022beans}.

For comparison, we also further pretrained ResNet and VGGish models with a supervised, multi-label classification objective on the combination of FSD50K, AudioSet's 20k, and {\tt bio} subsets. VGGSound follows its own label ontology and was not included in this configuration. We used a learning rate of $1.0 \times 10^{-4}$ with the Adam optimizer, and chose the best model based on the validation loss. Note that while our proposed \method~models do not rely on label information, these supervised ``topline'' models rely on annotations and are provided as a reference only.

\vspace{-0.5em}
\subsection{Fine-tuning and evaluation}
\vspace{-0.5em}

We ran a comprehensive evaluation on the BEANS benchmark~\cite{hagiwara2022beans}, which consists of the following datasets. The first five are used for classification and the next five for detection. We also included {\tt esc}~\cite{piczak2015dataset} (ESC-50) and {\tt sc}~\cite{warden2018speech} (SpeechCommands), two popular sound/speech classification datasets as a reference.

\vspace{-1em}
\begin{itemize}
\setlength\itemsep{-0.5em}
\item {\tt wtkn}~\cite{sayigh2016watkins} (The Watkins Marine Mammal Sound Database) is a database of marine mammal sounds which contains the recordings of 32 species.
\item {\tt bats}~\cite{prat2017annotated} contains Egyptian fruit bat calls. The target label is the emitter ID (10 individuals).
\item {\tt cbi}~\cite{cornell2020} is from the Cornell Bird Identification competition hoted on Kaggle containing 264 bird species.
\item {\tt dogs}~\cite{yin2004barking} contains dog barks recorded from 10 individual domestic dogs in different situations (disturbance, isolation, and play).
\item {\tt hbdb}~\cite{kiskin2021humbugdb} (HumBugDB) is a collection of wild and cultured mosquito wingbeat sounds. 
\item {\tt dcase}~\cite{morfi2021fewshot} is from DCASE 2021 Task 5: Few-shot Bioacoustic Event Detection It contains mammal and bird multi-species recordings with annotations. 
\item {\tt enab}~\cite{chronister2021annotated} (Eastern North American Birds) contains recordings of bird dawn chorus with annotations.
\item {\tt hiceas}~\cite{noaa2022hawaiian} is a dataset from the Hawaiian Islands Cetacean and Ecosystem Assessment Survey (HICEAS) in 2017. We used Minke whale ``boing'' vocalization annotations of the dataset.
\item {\tt rfcx}~\cite{lebien2020pipeline} is the dataset from Rainforest Connection (RFCx) with frog and bird sound.
\item {\tt gib}~\cite{dufourq2021automated} contains Hainan gibbon calls. The data were annotated with onset/offset times and call types. 
\end{itemize}
\vspace{-1em}

Each model was trained from scratch or fine-tuned on the training portion of each dataset using Adam with $\beta_1 = 0.9, \beta_2 = 0.999, \epsilon = 1.0\times10^{-8}$, and a batch size of 32 instances. We swept the learning rate over $1.0\times10^{-5}, 5.0\times10^{-5}, 1.0\times10^{-4}$ for 50 epochs, and picked the best model based on the validation metric, and evaluated it on the test split. We resampled the input at 16kHz for VGGish and \method, as these models are pretrained at this specific sampling rate. We used accuracy for evaluating classification, and mean average precision for evaluating detection tasks.

\vspace{-0.5em}
\subsection{Main results}
\vspace{-0.5em}

Table~\ref{table:results} shows the performance of all models, including the baselines (originally reported in~\cite{hagiwara2022beans}) and the toplines. Our model pretrained on the bioacoustics data (\method-{\tt bio}) outperformed all the baselines in all but two of the datasets we investigated. Overall, it was the best model and its performance even surpassed the supervised toplines (resnet50p-s, a supervised pretrained ResNet50, and VGGish models supervised on millions of YouTube videos).

Performance on some detection datasets (e.g., {\tt rfcx}, {\tt gibbons}) was lower than others. These datasets are challenging due to the sparsity of the vocalizations in the training data. \method~was the only model that performed competitively across all datasets. We expect modern regularization and data augmentation techniques, such as mixup~\cite{zhang2018mixup} and SpecAugment~\cite{park2019specaugment} to help improve the results further.

\vspace{-0.5em}
\subsection{Ablation on pretrainig data and model size}
\vspace{-0.5em}

\begin{table}[!t]
\begin{center}
{\small
\begin{tabular}{@{}lrrrr@{}}
\toprule
 Config.      & Classification    & Detection  & Auxiliary  & Total \\ \midrule
\method-{\tt core}   & 59.2       & 62.6      &  58.6        & 60.5   \\
\method-{\tt bio}    & 61.2       & 62.6      &  59.0        & 61.4   \\
\method-{\tt nonbio} & 61.8       & 61.7      & 58.8         & 61.3   \\
\method-{\tt all}    & 60.9       & 60.0      & 59.3         & 60.3  \\ \bottomrule
\end{tabular}
}
\end{center}
\vspace{-1.0em}
\caption{Relative T-scores on different dataset configurations}
\vspace{-1.0em}
\label{table:size}
\end{table}

Table~\ref{table:size} shows T-scores (normalized metrics within each dataset so that the average is 50 and the standard deviation is 10) for each pretraining configuration and task group. It shows that all configurations performed somewhat similarly, because we pretrained all models for the same number of steps, although \method-{\tt bio} outperforms all the other configurations with a small margin. This may indicate the effectiveness of choosing the pretraining data related to the downstream tasks, and extracting a small, diverse subset is the key to cost-effective learning~\cite{sorscher2022beyond}.

\begin{table}[!t]
\begin{center}
{\small
\begin{tabular}{@{}lrrrr@{}}
\toprule
 Config.      & Class.    & Detection  & Auxiliary  & Total \\ \midrule
\method-{\tt bio}-base    & 61.2       & 62.6      &  59.0        & 61.4   \\
\method-{\tt bio}-large   & 56.8       & 57.9      &  56.4        & 57.2   \\ \bottomrule
\end{tabular}
}
\end{center}
\vspace{-1.0em}
\caption{Relative T-scores on different model configurations}
\vspace{-1.0em}
\label{table:model}
\end{table}

We compared model size configurations (i.e., {\tt base} and {\tt large}) and their performance on each task in Table~\ref{table:model}. We found that large models consistently underperform their baseline counterparts in all configurations, including {\tt all}\footnote{We also tried data2vec~\cite{baevski2022data2vec}, another type of self-supervised method, although the preliminary results were not promising}. We posit that this is due to overfitting and an undertuning of model-specific hyperparameters. 

\vspace{-0.5em}
\subsection{Visualization of learned representations}
\vspace{-0.5em}

Finally, we show a t-SNE plot of pooled representation vectors produced for ESC-50 by the pretrained (not fine-tuned) \method-{\tt bio} model in Figure~\ref{fig:overview} (c). The plot shows that \method~was able to learn similar representations for the same and closely related animals (e.g., birds and crows), explaining its strong performance on many classification datasets.

\vspace{-0.5em}
\section{Conclusion}
\vspace{-0.5em}

In this paper, we proposed \method, a self-supervised, transformer-based audio representation model for encoding animal vocalizations, pretrained on a large amount of unnanotated audio datasets. Our model outperformed all baselines and even supervised toplines on a diverse set of bioacoustics tasks. 
We believe that our models would benefit from further scale in terms of both data and model size, as well as modern regularization and data augmentation techniques, which we leave for future work\footnote{The author would like to thank Benjamin Hoffman, Felix Effenberger, Jen-Yu Liu, and Maddie Cusimano for their valuable feedback.}.

\setlength{\bibitemsep}{0\baselineskip}

\bibliographystyle{IEEEbib}
\bibliography{refs_shorter}

\end{document}